\documentclass[pra,aps,twocolumn,amsmath,amssymb,superscriptaddress,color,showpacs]{revtex4}

\usepackage{multirow}
\usepackage{amssymb}
\usepackage{amsmath}
\usepackage{graphicx}
\usepackage{amsfonts}
\usepackage{color}
\usepackage{bm}
\usepackage{subfigure}
\usepackage{xspace}

\usepackage{ulem}

\usepackage{soul,xcolor}
\setstcolor{blue}

\newcommand{\TG}{TG\xspace}

\begin{document}

\title{Breakdown of  scale invariance in the vicinity of the Tonks-Girardeau limit}
\author{Z.\ D.\ Zhang}
\email{zhedong.zhang@stonybrook.edu}
\affiliation{Department of Physics and Astronomy, SUNY Stony Brook, NY 11794  USA}
\author{G.\ E.\ Astrakharchik}
\affiliation{Departament de F\'\i sica i Enginyeria Nuclear, Campus Nord B4-B5,
Universitat Polit\`{e}cnica de Catalunya, E-08034 Barcelona, Spain}
\author{D.\ C.\ Aveline}
\affiliation{Jet Propulsion Laboratory, California Institute of Technology, CA 91109 USA }
%
\author{S.\ Choi}
\affiliation{Department of Physics, University of Massachusetts Boston, Boston Massachusetts 02125, USA}
\author{H. Perrin}
\affiliation{Laboratoire de physique des lasers, CNRS, Universit\'{e} Paris 13,
Sorbonne Paris Cit\'{e}, 99 avenue J.-B. Cl\'{e}ment, F-93430 Villetaneuse, France}
\author{T.\ H.\ Bergeman}
\affiliation{Department of Physics and Astronomy, SUNY Stony Brook, NY 11794  USA}
\author{M.\ Olshanii}
\affiliation{Department of Physics, University of Massachusetts Boston, Boston Massachusetts 02125, USA}
\date{\today}

\begin{abstract}
In this article, we consider the monopole excitations of the harmonically trapped Bose gas
in the vicinity of the Tonks-Girardeau limit. 
Using Girardeau's Fermi-Bose duality and subsequently
an effective fermion-fermion odd-wave interaction, we obtain the dominant correction
to the scale-invariance-protected value of the excitation frequency, for microscopically small excitation amplitudes.
We produce a series of diffusion Monte Carlo results that confirm our analytic prediction for three particles.
And less expectedly, our result stands in excellent agreement with the result of a hydrodynamic simulation (with the Lieb-Liniger equation of state as an input) of the microscopically large but macroscopically 
small excitations. We also show that the frequency we obtain coincides with the upper bound derived 
by Menotti and Stringari using sum rules.  
Surprisingly, however, we found that the usually successful hydrodynamic perturbation theory predicts a shift that
is $9/4$ higher than its \textit{ab initio} numerical counterpart.
We conjecture that the sharp boundary of the cloud in local density approximation---characterized
by an infinite density gradient---renders the
perturbation inapplicable. All our results also directly apply to the 3D $p$-wave-interacting 
waveguide-confined confined fermions.
\end{abstract}


\maketitle

%
%
\section{Introduction}
In scale-invariant systems, a state at one density can be expressed through a state at another, via trivial rescaling of space.
Scale invariance is always associated with an inability of the interaction potential to introduce a distinct length scale.
Several examples emerged recently in the physics of quantum gases. 
In three dimensions, the $\delta$-interaction with infinite coupling strength, even when properly regularized, ensures the scale invariance of the unitary gases
\cite{ohara2002_2179,gelm2003_011401,bourdel2003_020402,gupta2003_1723,regal2013_230404}. In 2D, the unregularized $\delta$-potential,
for any coupling constant, induces the scale invariance of
two-dimensional Bose \cite{hung2011_236,chevy2001_250402} and spin-$1/2$ Fermi
\cite{vogt2012_070404} gases at the classical field level, which is however broken by
quantization \cite{olshanii2010_095302,taylor2012_135301,hofmann2012_185303}. Finally, in 1D, we have the Tonks-Girardeau (\TG) gas
\cite{girardeau1960_516,kinoshita2004_1125,paredes2004_277}---a one-dimensional quantum Bose gas with an infinite strength $\delta$-interaction---which
is the subject of this article.

Scale invariance enables a robust frequency gauge: when a scale-invariant gas is placed in a symmetric harmonic trap of frequency
$\omega$ and
a monopole oscillation is induced, the signal shows neither damping nor amplitude-dependent frequency shifts---its frequency is fixed to
$2 \omega$, for all scale invariant systems and for all spatial dimensions \cite{pitaevskii1997}. As a consequence,
monopole excitations in scale-invariant systems are very sensitive to changes in the equation of state \cite{silva2006_0607491},
whether produced
by a quantum anomaly \cite{olshanii2010_095302,taylor2012_135301,hofmann2012_185303}, by an influence of the confining dimension
\cite{merloti2013_13111028}, or just by a small shift in the coupling constant away from the scale-invariant point.

In this article, we study the effect of a small deviation from the \TG point
on the frequency of the
monopole excitations
of a one-dimensional harmonically trapped Bose gas \cite{menotti2002_043610,haller2009_1224},
both for microscopically small 
and for microscopically large but macroscopically small
excitation amplitudes. Our results also directly apply to the 3D $p$-wave-interacting waveguide-confined 
fermions \cite{gunter2005_230401}, thanks to the mapping by Granger and Blume
\cite{granger2004_133202}.

%
%
\section{Bosonic Hamiltonian of interest and its effective fermionic counterpart}
Our object of study is the system of $N$ bosons of mass $m$ with contact interaction in a
1D harmonic trap of frequency $\omega$. In the second quantized form, the Hamiltonian reads
\begin{align}
\begin{split}
&
\hat{H}_{B} =
\int_{-\infty}^{+\infty}\!dx\,
\left\{
\frac{\hbar^{2}}{2m}
(\partial_x \hat{\Psi}_{B}^{\dagger})(\partial_x \hat{\Psi}_{B}^{})
\qquad
\right.
\\[0.16cm]
&
\qquad\qquad
\left.
+
\frac{m\omega^2}{2} x^2 \hat{\Psi}_{B}^{\dagger}\hat{\Psi}_{B}^{}
+\frac{g_{\mbox{\scriptsize 1D}}}{2}
\hat{\Psi}_{B}^{\dagger}
\hat{\Psi}_{B}^{\dagger}
\hat{\Psi}_{B}^{}
\hat{\Psi}_{B}^{}
\right\}
\,.
\end{split}
\label{1m}
\end{align}
Here $g_{\mbox{\scriptsize 1D}}$ is the one-dimensional coupling
constant,
and $\hat{\Psi}_{B}(x)$ is the bosonic quantum field.

In this article, we will be interested in the monopole excitations
in the vicinity of the \TG limit, $g_{\mbox{\scriptsize 1D}} \to \infty$.
This regime is both easy and difficult to work in.
On one hand, right at the limit, the model maps to free fermions, via the
Fermi-Bose map by Girardeau \cite{girardeau1960_516}. On the other hand, away from the
limit---even remaining infinitesimally close to it---there exist conceptual difficulties in
interpreting the resulting system as one governed by a Hamiltonian for free fermions plus a small correction
\cite{remark_on_impossible_potentials_bis}.

Nevertheless, an effective fermionic Hamiltonian,
%
%
\begin{align}
\begin{split}
&
\hat{H}_{F\mbox{\scriptsize, eff.}} =
\int_{-\infty}^{+\infty}\!dx\,
\left\{
\frac{\hbar^{2}}{2m}(\partial_x\hat{\Psi}_{F}^{\dagger})(\partial_x \hat{\Psi}_{F}^{})
\right.
\\[0.16cm]
&
\left.
+
\frac{m\omega^2}{2} x^2 \hat{\Psi}_{F}^{\dagger}\hat{\Psi}_{F}^{}
-
\frac{2\hbar^4}{m^2 g_{\mbox{\scriptsize 1D}}}
(\partial_x\hat{\Psi}_{F}^{\dagger})
\hat{\Psi}_{F}^{\dagger}
\hat{\Psi}_{F}^{}
(\partial_x\hat{\Psi}_{F}^{})
\right\}
\end{split}
\label{4m}
\,,
\end{align}
\cite{girardeau2004_023608}
can be proven \cite{girardeau2003_0309396,diptiman2003_7517} to produce the
correct eigenspectrum if used as the kernel of a variational energy functional. One can further show that in this case,
the first order of the perturbation theory---with the quartic term in (\ref{4m}) as a perturbation---produces the correct
$1/g_{\mbox{\scriptsize 1D}}$ correction to the eigenenergies.
Here, $\hat{\Psi}_{F}(x)$
is the fermionic quantum field.
%
%

%
%
\section{Frequency of the monopole excitation of a microscopically small amplitude}
The fermionic field in Eq.~(2) can be expanded 
onto a series over the eigenstates of the harmonic trap:
$\hat{\Psi}_{F}(x)=\sum_{n} \hat{b}_n \varphi_n (x),\quad
\varphi_n (x)=\left[1/(2^n n! \sqrt{\pi} \ell)\right]^{1/2}
e^{-x^2/(2 \ell^2)} H_n (x/\ell)$
where $H_n(\xi)$ is the
${n}$-th Hermite polynomial, and $\ell \equiv \sqrt{\hbar/(m\omega)}$. 
The operator $\hat{b}_n^{}$
is the fermionic annihilation operator that removes one particle from the $n$-th eigenstate. The 
operators $\hat{b}_n^{}$ obey the standard fermionic 
commutation relations and the Hamiltonian in Fock space is of the form
\begin{equation}
\begin{split}
\hat{H}_{F\mbox{\scriptsize, eff.}} & =\frac{N}{2}\hbar\omega+\sum_{n=0}^{\infty}n\hbar\omega \hat{b}_n^{\dagger}\hat{b}_n\\
&  \quad -\frac{\hbar^4}{m^2g_{\mbox{\scriptsize 1D}}\ell^3}\underset{n+m+k+l=\textup{even}}{\sum_{n<m}^{\infty}\sum_{k<l}^{\infty}}\Omega_{kl}^{nm}\hat{b}_n^{\dagger}\hat{b}_m^{\dagger}\hat{b}_k\hat{b}_l
\end{split}
\label{5tris}
\end{equation}
where $\Omega_{kl}^{nm}=2\int_{-\infty}^{\infty}\textup{d}\xi
\left(\varphi'_n\varphi_m-\varphi_n\varphi'_m\right)\left(\varphi'_l\varphi_k-\varphi_l\varphi'_k\right)$. The ground state of whole system is
\begin{equation}
\begin{split}
|\Psi_0\rangle=\left(\prod_{n=0}^{N-1}\hat{b}_n^{\dagger}\right)|vac\rangle
\end{split}
\label{6tris}
\end{equation}
where $|vac\rangle$ stands for the vacuum with no particle at all. The energy correction of ground state is analyzed in Appendix A, where we will show our result recovers the formula in Ref.~\cite{paraah2010_065603}.

Now we will come to the $2^{\textup{nd}}$ excitations. The unperturbed manifold of energy $E_0^{(0)}+2\hbar\omega$ is of two-fold degeneracy. The set of unperturbed eigenstates is $\left\{
|\Psi_{2a}\rangle=\hat{b}_{N+1}^{\dagger}
\hat{b}_{N-1}^{}|\Psi_0\rangle,\,
|\Psi_{2b}\rangle=\hat{b}_N^{\dagger}
\hat{b}_{N-2}^{}|\Psi_0\rangle
\right\}$. Based on the perturbation theory, corrections to the energies are represented by the spectrum of the $2\times 2$ matrix of the perturbation term in the space spanned by the members of the manifold
\begin{equation}
\begin{split}
\hat{\mathcal{V}}=\frac{\hbar^4}{m^2 g_{1D} \ell^3}\begin{pmatrix}
                                                     I_N^{(2a)} & \Omega_N\\[0.2cm]
                                                     \Omega_N & I_N^{(2b)}\\
                                                   \end{pmatrix}
\end{split}
\label{7tris}
\end{equation}
with $\Omega_N \equiv \Omega_{N-1,N}^{N-2,N+1}$, 
$I_N^{(2a)} = \sum_{m=1}^{N+1(a)}\sum_{n=0}^{m-1}\upsilon_{nm}$, and 
$I_N^{(2b)} =\sum_{m=1}^{N(b)}\sum_{n=0}^{m-1}\upsilon_{nm}$. Hence the transition frequencies for microscopically small amplitude read
\begin{equation}
\begin{split}
& \hbar\omega_{2\pm,\,0}=2\hbar\omega+ \frac{1}{2} \frac{\hbar^4}{m^2 g_{\mbox{\scriptsize 1D}} \ell^3}
\bigg[I_N^{(2a)}+I_N^{(2b)}-2 I_N^{(0)}\\[0.1cm]
& \quad\pm\sqrt{\big(I_N^{(2a)}-I_N^{(2b)}\big)^2+4\Omega_N^2}\ \bigg] + {\cal O}(1/(g_{\mbox{\scriptsize 1D}})^2)
\end{split}
\label{8tris}
\end{equation}
where $I_N^{(0)}\equiv\sum\limits_{m=1}^{N-1}\sum\limits_{n=0}^{m-1}\upsilon_{nm}$ and $\upsilon_{nm}$ takes the form of

\begin{equation}
\begin{split}
\upsilon_{nm}=\sqrt{\frac{2}{\pi^3}}& \frac{(m-n)^2
\Gamma\left(m-\frac{1}{2}\right)}{\Gamma\left(m+1\right)}
\frac{\Gamma\left(n-\frac{1}{2}\right)}{\Gamma\left(n+1\right)}\\[0.2cm]
& \times{}_3 F_2\left[\begin{matrix}\frac{3}{2},-n,-m\\[0.08cm]
\frac{3}{2}-n,\frac{3}{2}-m\end{matrix};1\right]
\end{split}
\label{16b}
\end{equation}
 After a lengthy but straightforward calculation (see the Appendices),
we obtain the analytical form of $1/g_{\mbox{\scriptsize 1D}}$
corrections to the relevant transition frequencies:
\begin{equation}
\begin{aligned}
\hbar\omega_{2+,\,0} \equiv 2\, \hbar\Omega_{\mbox{\scriptsize D}} =
\left(
  2 + {\cal O}(\frac{1}{\gamma_{0}^2(N)})
\right) \hbar\omega
\end{aligned}
\label{12}
\end{equation}
\begin{align}
&
\hbar\omega_{2-,0} \equiv \hbar\Omega_{\mbox{\scriptsize M}} =
\left(
  2
  - \frac{6}{\sqrt{\pi}}
  \frac{\sqrt{N}\Gamma\left(N-\frac{5}{2}\right)
  \Gamma\left(N+\frac{1}{2}\right)}{\Gamma\left(N\right)\Gamma\left(N+2\right)}
  \right.
  \nonumber
  \\[0.2cm]
  &
  \left.
  \times {}_3 F_2\left[\begin{matrix}\frac{3}{2},1-N,-N\\[0.08cm]
  \frac{7}{2}-N,\frac{1}{2}-N\end{matrix};1\right]
  \frac{1}{\gamma_{0}(N)}
  + {\cal O}(\frac{1}{\gamma_{0}^2(N)})
\right) \hbar\omega
\,,
\label{13m}
\end{align}
where $\hbar\omega_{2\pm,\,0} = E_{2\pm} - E_{0}$, are the transition frequencies,
$E_{0}$ is the ground state energy, and $E_{2\pm}$ are the energies of the
states that, in the strict \TG limit, form a two-fold degenerate manifold,
$2 \hbar\omega$ above the ground state.
The effective Lieb-Liniger parameter $\gamma_0(N)\equiv (m g_{\mbox{\scriptsize 1D}})/(n_{\mbox{\scriptsize TF}}\hbar^2)$ \cite{lieb1963_1605} uses the \TG (i.e. $g_{\mbox{\scriptsize 1D}} \to \infty$) density in the center of the trap,
$n_{\mbox{\scriptsize TF}} \equiv (\sqrt{2}/\pi) \sqrt{N} \sqrt{m\omega/\hbar}$,
instead of the true density. Here, ${}_3 F_2\left[a_1,\,a_2,\,a_3;\,
b_1,\,b_2;\,z\right]$ is the generalized hypergeometric function of order $(3,\,2)$.

The interpretation of the $2\pm$ eigenstates can be inferred from the corresponding transition frequencies.

The first one ($2+$) is the second state of an infinite $\hbar\omega$-spaced ``dipole'' ladder: coherent wave
packets formed out of the members of the ladder represent finite amplitude dipole excitations; their frequency
$\Omega_{\mbox{\scriptsize D}}$ is equal to the frequency of the trap \textit{exactly}, interactions notwithstanding
\cite{kohn1961_1242}. The first state of the ladder is analyzed in Appendix B. The zeroth state is the
ground state.

The second eigenstate ($2-$) in the $E_0^{(0)} + 2\hbar\omega$ manifold, is the first (ground state being the zeroth)
step in the ``monopole'' ladder, that corresponds to the breathing excitations of frequency $\Omega_{\mbox{\scriptsize M}}$.
In the noninteracting case, the ladder (exactly $2\hbar\omega$-spaced) can be obtained by a recurring application
of the creation operator $\hat{L}_{+}$ of an appropriate
$SO(2,1)$ group to the ground state \cite{pitaevskii1997}. The excitation dynamics
consists of a periodic scaling transformation of frequency $2\omega$. The existence of this structure is a direct
consequence of the scale invariance of the \TG gas
and its free-fermionic counterpart in the harmonic potential.

A deviation from the \TG limit (and the corresponding fermion-fermion interactions (\ref{4m}))
\textit{breaks the scale invariance weakly}. The goal of this article is to
assess the impact that this effect has on the excitations of a microscopic amplitude and compare it to the 
corresponding predictions for the microscopically large but macroscopically small excitations.

As far as the microscopic amplitude excitations are concerned, our program is already fulfilled. Indeed, a linear combination of the ground
state and a small admixture of the state $2-$ \textit{is} is already a small amplitude monopole excitation.  Its frequency is given
by the formula (\ref{13m}) that \textit{constitutes the central result of this article}.

%
%
\section{Comparison to the other few-body results}
We verified that for two atoms ($N=2$), the formula (\ref{13m}) for the frequency of the small amplitude monopole
excitations coincides with the known exact results \cite{busch1998_549}.

In the three-body case ($N=3$) we perform a Diffusion Monte Carlo simulation of the imaginary time evolution
and extract the $\omega_{2-,\,0}$ transition frequency from the inverse Laplace transform components of the
imaginary-time dynamic structure factor.

\begin{figure}
\centering\includegraphics[scale=.6]{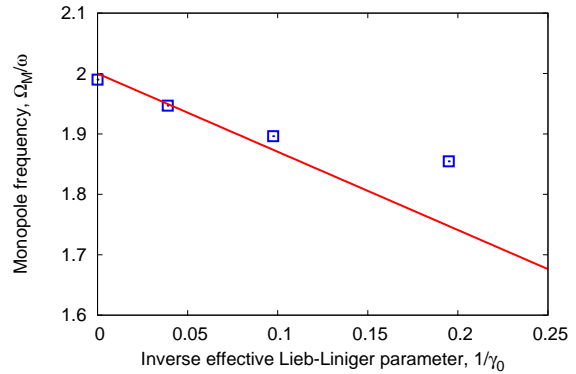}
\caption{
(color online).
The frequency of a small amplitude monopole excitation for $N = 3$ one-dimensional bosons in a harmonic trap,
as a function of the inverse of the effective Lieb-Liniger parameter $\gamma_{0}$ (see text).
Solid line (red online):
the prediction of the formula (\ref{13m}) for the first two terms of the expansion of frequency in powers
of $1/\gamma_{0}$. Open squares (blue online): the ab initio Diffusion Monte Carlo simulation.
       }
\label{Fig.3}
\end{figure}

\begin{figure}
\centering\includegraphics[scale=.6]{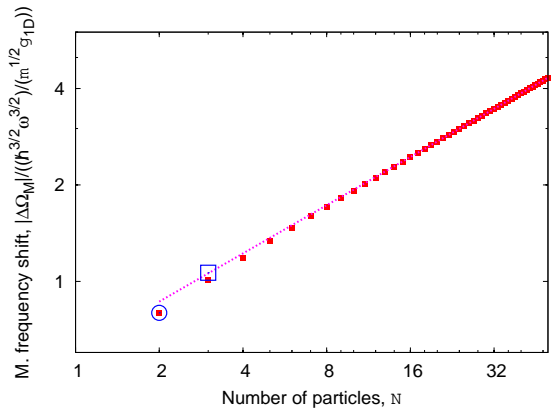}
\caption{(color online).
The magnitude of the dominant correction, in a power-series expansion in $1/\gamma_{0}$, to the result predicted by the scale invariance, $\Omega_{\mbox{\scriptsize M}} = 2 \omega$.
Filled squares (red online): the analytic formula (\ref{13m}). Open circle (blue online): the exact nonperturbative solution for $N=2$
\cite{busch1998_549}.
Open square (blue online): the Diffusion Monte Carlo simulation for $N=3$. Dotted line (purple online): the $N \gg 1$ limit of the
sum-rule prediction \cite{menotti2002_043610, astrakharchik2005_063620} (also Eq.~(\ref{14m})).
        }
\label{Fig.2}
\end{figure}

In Fig.~\ref{Fig.3} we compare our (non-perturbative) numerical three-body results with the perturbative
prediction (\ref{13m}). In Fig.~\ref{Fig.2}, the dominant corrections to the monopole frequency for
$N=2$ and $N=3$, extracted from the non-perturbative data, are also compared to formula (\ref{13m}).

%
%
\section{Large-$N$ asymptotics and a comparison with the sum-rule predictions}
\begin{figure}
\centering\includegraphics[scale=.6]{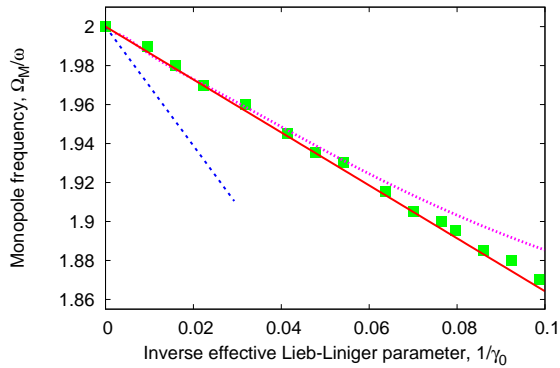}
\caption{
(color online).	
The frequency of both the microscopically small and microscopically large but macroscopically small monopole excitations, 
in the limit of $N \gg 1$.
Solid line (red online): the formula (\ref{13m}). Dotted line (purple online): the sum-rule upper bound \cite{menotti2002_043610}
(courtesy of Chiara Menotti and Sandro Stringari).
Filled squares (green online): the numerically exact hydrodynamic simulation of the motion 
of a macroscopic motion of small amplitude. Dashed line (blue online):
the hydrodynamic perturbation theory for the latter. For definition of $\gamma_{0}$, see text. 	
        }
\label{Fig.1}
\end{figure}

The frequency of the monopole excitations of a microscopically small amplitude can be bounded from above using the sum rules
\cite{menotti2002_043610} (see Fig.~\ref{Fig.1}). The order $1/g_{\mbox{\scriptsize 1D}}$ correction
to this bound can also be computed analytically, for large atom numbers \cite{astrakharchik2005_063620}:
\begin{align}
\hbar \Omega_{\mbox{\scriptsize M}}
\stackrel{N\gg 1}{=}
\left(
  2 - \frac{64}{15 \pi} \frac{1}{\gamma_{0}(N)} + {\cal O}(\frac{1}{\gamma_{0}^2(N)})
\right) \hbar\omega
\label{14m}
\end{align}
We conjecture that the upper bound (\ref{14m}) actually equals the exact prediction (\ref{13m}) in the limit
of large $N$. To test this conjecture, we multiplied the $\frac{1}{\gamma_{0}(N)}$-term in the
the bound (\ref{14m}) by $A N^{\sigma}$, with $A$ and $\sigma$
being free parameters to be used to fit the $\frac{1}{\gamma_{0}(N)}$ term in the series
(\ref{13m}). Indeed we found the values that
support our conjecture, namely $A = 1.000$ and $\sigma = 0.0003$. 

%

%
%
\section{Comparison to the frequencies of the excitations of a microscopically large but macroscopically small amplitude}
The monopole frequencies obtained above correspond to excitations
of  \textit{microscopically small amplitude}: there
the many-body energy of the excited atomic cloud is only a few one-body harmonic quanta above the ground state energy.
\textit{A priori} it is not obvious if the microscopic predictions
will remain valid
for \textit{microscopically large but macroscopically small} excitations, whose 
\textit{spatial amplitude is smaller than but comparable to the size of the cloud}.

To compare the two frequencies, we investigate the time dynamics 
using the hydrodynamic equations (see e.g. Eqs.~(1) and (2) of Ref.\ \cite{pitaevskii1998_4541}).
We use the well-known thermodynamic limit for the dependence of the zero-temperature chemical potential $\mu(n)$ on the one-dimensional
particle density $n$, for a uniform one-dimensional $\delta$-interacting Bose gas; this equation of state was
obtained by Lieb and Liniger, using Bethe Ansatz \cite{lieb1963_1605}.
We propagate the hydrodynamic equations numerically. To excite the monopole mode,
we quench the trapping frequency. Fig.~\ref{Fig.1} shows a good agreement with the large-$N$ asymptotics for
the frequency of the microscopically small excitations (\ref{14m}).

In order to obtain an analytic
expression for the frequency shift, we apply the perturbation theory developed by Pitaevskii and Stringari in
Ref.\ \cite{pitaevskii1998_4541} (with more technical details worked out in Ref.\ \cite{merloti2013_13111028})
for the purpose of computing an analytic expression for the dominant beyond-mean-field correction to the monopole frequency of a BEC.
Here we use the \TG
equation of state (EoS), $\mu_{0}(n)=(\pi^2\hbar^2/2m) n^2$ as the unperturbed EoS, and the first-order (in $\gamma(n)^{-1}$) correction to the EoS,
$\Delta\mu(n) = (8\pi^2\hbar^2/3m) n^2 \gamma(n)^{-1}$, as a perturbation. The function
$\gamma(n) \equiv (m g_{\mbox{\scriptsize 1D}})/(n\hbar^2)$ is the so-called
Lieb-Liniger parameter \cite{lieb1963_1605}. While most of the outlined steps of the study in \cite{pitaevskii1998_4541,merloti2013_13111028} are universally applicable
to any EoS, the boundary conditions
for the density mode functions $\delta n(z)$ at the edge of the atomic cloud $|z| = R_{\mbox{\scriptsize TF}}$ are typically dictated by the specific physical properties of the system at hand. (Here $R_{\mbox{\scriptsize TF}}$ is the Thomas-Fermi radius.)
In the \TG case, with or without further beyond-the-\TG corrections to the EoS, those are given by
\begin{align}
&
\delta n(z) = A (R_{\mbox{\scriptsize TF}}-|r|)^{-1/2} + B + {\cal O}((R_{\mbox{\scriptsize TF}}-|r|)^{1/2})
\nonumber
\\
&
B = 0
\label{bc}
\,.
\end{align}
Indeed, following
the analysis developed in Ref.\ \cite{merloti2013_13111028}, one can show (i) that the first
two terms in (\ref{bc}) correspond to the near-edge asymptotic of the two linearly independent
solutions of the mode equation, and (ii) that when rewritten
in Lagrange form \cite{landau_hydrodynamics},
the solutions that violate condition (\ref{bc}) lead to the appearance
of crossing particle trajectories, incompatible with hydrodynamics.

To our surprise, we found that the macroscopic perturbation theory leads to a frequency shift that is $9/4$ times greater in magnitude than
its microscopic counterpart Eq.~(\ref{14m}) (see Fig.~\ref{Fig.1}). %
%
This is definitely an artifact of the perturbative treatment of the macroscopic theory rather than of the macroscopic theory per se. Indeed, our macroscopic nonperturbative numerical results are consistent
with the microscopic theory. We attribute the failure of the perturbation theory to the divergence of the spatial derivative of the
steady-state density at the edge of the cloud: in a monopole excitation this will lead to an infinite time derivative of the
density itself, possibly invalidating the perturbation theory.

In the same plot,  we also present the sum-rule bound \cite{menotti2002_043610}. At weak fermion-fermion interactions,
it reproduces well the perturbative prediction Eq.~(\ref{14m}).

%
%
\section{Conclusion and outlook}
In this article, we obtained an analytic expression, Eq.~(\ref{13m}) for
the leading behavior of the deviation
of the frequency of the microscopically small monopole
excitations of a strongly-interacting one-dimensional Bose gas from the value predicted by the scale invariance in the \TG limit.

We further compare this prediction with (a) the known non-perturbative analytic expressions for two atoms \cite{busch1998_549} and to
(b) the Diffusion Monte Carlo predictions for three atoms. For large numbers of atoms, the prediction in Eq.~(\ref{13m})
stands in excellent agreement with (c$\mbox{)}$ the sum-rule bound (\ref{14m}) \cite{menotti2002_043610, astrakharchik2005_063620}.
It was not {\it a priori} 
obvious to us if our formula will also apply to
\textit{microscopically large but macroscopically small} 
excitations: they correspond to a large number of atoms (still covered by the formula (\ref{13m})),
and have a macroscopic magnitude (that is formally beyond the scope of Eq.~(\ref{13m})).
We found that (d) the numerically propagated hydrodynamic equations produce
the same leading-order frequency correction as the large-$N$ limit of Eq.~(\ref{13m}).
Finally, we find that (e) the hydrodynamic perturbation theory,
which was so successful in predicting the beyond-mean-field corrections to the monopole
frequency in both the three-dimensional \cite{pitaevskii1998_4541} and the two-dimensional \cite{merloti2013_13111028} Bose gases, fails to predict
the analogous beyond-\TG correction in our case: the hydrodynamic perturbative prediction turns out to be approximately $9/4$
higher than the ab initio
%
%
numerical value it was designed to approximate. We  conjecture that the sharp
boundary of the \TG cloud, characterized by an infinite density gradient, renders the perturbation theory
inapplicable.

Experimentally, the monopole excitation frequency of the Lieb-Liniger gas has been already studied, in
Ref.\ \cite{haller2009_1224}. In the range of
parameters our article is devoted to, the beyond-scale-invariance shifts are too small to be reliably compared with the experimental data.
However, we plan to extend our study of the frequency of microscopically large but macroscopically small 
monopole excitations to the whole range of the interaction strengths.
One can already observe that in the intermediate range, the experimental frequencies \cite{haller2009_1224} depart from the sum-rule upper bound
\cite{menotti2002_043610}. It appears to be of interest to verify that the numerically propagated hydrodynamic
equations can reproduce the experimental points.

A study of the \textit{finite amplitude} beyond-\TG corrections to the monopole frequency may be of interest.
Another possible direction
is computing the higher orders of the perturbation theory for the frequency correction. This step
is challenging, however: the odd-wave fermion-fermion
interaction potential in Eq.~(\ref{4m}) cannot be used as such, and it requires a prior
regularization \cite{remark_on_impossible_potentials_bis}, similar to the
Fermi-Huang regularization of the three-dimensional $\delta$-potential.

Results of our work directly apply to another system: the spin-polarized $p$-wave-interacting fermions in a wave-guide
\cite{gunter2005_230401}. The mapping between this
system and the Lieb-Liniger gas of $\delta $-interacting bosons is provided by Granger and Blume,
in the final formula of Ref.\ \cite{granger2004_133202}.
In the case of $^{40}K$ atoms, the $p$-wave scattering volume $V_{p}$ can be controlled at will, using an accessible $l=1$, $m_{l}=0$
Feshbach resonance at $198.8\, \mbox{G}$ \cite{regal2003_053201}. When atoms are confined to a one-dimensional harmonic
waveguide, the position of the resonance is further shifted \cite{gunter2005_230401} due to the presence of
a confinement-induced resonance (CIR) \cite{granger2004_133202}. For example,
an ensemble of $N=500$ $^{40}K$ atoms, transversally frozen to a harmonic waveguide of
a confining frequency of $2\pi \times 25\,\mbox{kHz}$ and longitudinally
trapped by a  harmonic potential of frequency $2\pi \times 5\, \mbox{Hz}$,
will show a $\omega_{2-,\,0}/\omega - 2 = -4.7\%$ $p$-wave-interaction-induced shift of the monopole frequency, for the $p$-wave
scattering volume of $V_{p} = -(1000\, a_{B})^3$ (with the  CIR value situated at
$(V_{p})_{\mbox{\scriptsize{CIR}}} = -(2004\, a_{B})^3$). To relate this value of the scattering volume to the detuning from the
Feshbach resonance, note that the above value of the scattering volume would correspond to a binding energy of the three-dimensional
$p$-wave dimers \cite{granger2004_133202,gaebler2007_200403} of
$E_{\mbox{\scriptsize dimer}} \equiv -\hbar^2/2\tilde{\mu} (V_{p})^{2/3} = h \times 92.3 \mbox{kHz}$; the later value occurs if the magnetic
field is detuned by $0.49\,\mbox{G}$ below the $l=1$, $m_{l}=0$ resonance (see a the caption to Fig.\ 2 in \cite{gaebler2007_200403}
for the slope of the binding energy vs.\ magnetic field curve). In general, the slope of the dimer energy as a function
of the magnetic field is measured to be $h \times 188 \pm 2\, \mbox{kHz/G}$. Here, $\tilde{\mu} = m/2$ is the reduced mass, and $h= 2\pi \times \hbar$
is Plank's constant.

\section{Acknowledgments}
We acknowledge support from the Institut Francilien de Recherche sur les Atomes Froids (IFRAF). GEA acknowledges financial support from by MEC (Spain) through the Ramon y Cajal fellowship program, DGI (Spain) Grant No.~FIS2011-25275 and Generalitat de Catalunya Grant No. 2009SGR-1003.
ZDZ and TB were supported by the NSF grant (PHY-0968905).
MO was supported by grants from the Office of Naval Research
(N00014-12-1-0400) and the NSF grant (PHY-1019197).
Laboratoire de physique des lasers is UMR 7538 of CNRS and Paris 13 University.


\begin{thebibliography}{}
\bibitem{ohara2002_2179}
K.\ M.\ O'Hara, S.\ L.\ Hemmer, M.\ E.\ Gehm, S.\ R.\ Granade, and J.\ E.\ Thomas, Science \textbf{298}, 2179 (2002).
\bibitem{gelm2003_011401}
M.\ E.\ Gehm, S.\ L.\ Hemmer, S.\ R.\ Granade, K.\ M.\ O'Hara, and J.\ E.\ Thomas, Phys. Rev. A \textbf{68}, 011401 (2003).
\bibitem{bourdel2003_020402}
T.\ Bourdel, J.\ Cubizolles, L.\ Khaykovich, K.\ M.\ F.\ Magalhaes, S.\ J.\ J.\ M.\ F.\ Kokkelmans, G.\ V.\ Shlyapnikov, and C.\ Salomon, Phys. Rev. Lett. \textbf{91}, 020402 (2003).
\bibitem{gupta2003_1723}
S.\ Gupta, Z.\ Hadzibabic, M.\ W.\ Zwierlein, C.\ A.\ Stan, K.\ Dieckmann, C.\ H.\ Schunck, E.\ G.\ M.\ van Kempen, B.\ J.\ Verhaar, and W.\ Ketterle, Science \textbf{300}, 1723 (2003).
\bibitem{regal2013_230404}
C.\ A.\ Regal and D.\ S.\ Jin, Phys. Rev. Lett. \textbf{90}, 230404 (2003)
\bibitem{hung2011_236}
C.\ L.\ Hung, X.\ B.\ Zhang, N.\ Gemelke, and C.\ Chin, Nature \textbf{470}, 236 (2011).
\bibitem{chevy2001_250402}
F.\ Chevy, V\. Bretin, P.\ Rosenbusch, K.\ W.\ Madison, and J.\ Dalibard, Phys. Rev. Lett. \textbf{88}, 250402 (2002).
\bibitem{vogt2012_070404}
E.\ Vogt, M.\ Feld, B.\ Frohlich, D.\ Pertot, M.\ Koschorreck, and M.\ Kohl, Phys. Rev. Lett. \textbf{108}, 070404 (2012).
\bibitem{olshanii2010_095302}
M.\ Olshanii, H.\ Perrin, and V.\ Lorent, Phys. Rev. Lett. \textbf{105}, 095302 (2010).
\bibitem{taylor2012_135301}
E.\ Taylor and M.\ Randeria, Phys. Rev. Lett. \textbf{109}, 135301 (2012).
\bibitem{hofmann2012_185303}
J.\ Hofmann, Phys. Rev. Lett. \textbf{108}, 185303 (2012).
\bibitem{girardeau1960_516}
M.\ Girardeau, J. Math. Phys. \textbf{1}, 516 (1960).
\bibitem{kinoshita2004_1125}
T.\ Kinoshita, T.\ Wenger, and D.\ S.\ Weiss, Science \textbf{305}, 1125 (2004).
\bibitem{paredes2004_277}
B.\ Paredes, A.\ Widera, V.\ Murg, O.\ Mandel, S.\ F$\ddot{\textup{o}}$lling, I.\ Cirac, G.\ V.\ Shlyapnikov, T.\ W.\ H$\ddot{\textup{a}}$nsch, and I.\ Bloch, Nature \textbf{429}, 277 (2004).
\bibitem{pitaevskii1997}
L.\ P.\ Pitaevskii and A.\ Rosch, Phys. Rev. A \textbf{55}, R853 (1997).
\bibitem{silva2006_0607491}
T.\ N.\ D.\ Silva and E.\ J.\ Mueller, Preprint at arXiv:cond-mat/0607491 (2006).
\bibitem{merloti2013_13111028}
K.\ Merloti, R.\ Dubessy, L.\ Longchambon, M.\ Olshanii, and H.\ Perrin, Phys. Rev. A \textbf{88}, 061603(R) (2013)
\bibitem{menotti2002_043610}
C.\ Menotti and S.\ Stringari, Phys. Rev. A \textbf{66}, 043610 (2002).
\bibitem{haller2009_1224}
E.\ Haller, M.\ Gustavsson, M.\ Mark, J.\ Danzl, R.\ Hart, G.\ Pupillo, and H.-C.\ N$\ddot{\textup{a}}$gerl, Science \textbf{325}, 1224 (2009).
\bibitem{gunter2005_230401}
K.\ Gunter, T.\ Stoferle, H.\ Moritz, M.\ Kohl, and T.\ Esslinger, Phys. Rev. Lett. \textbf{95}, 230401 (2005).
\bibitem{granger2004_133202}
B.\ E.\ Granger and D.\ Blume, Phys. Rev. Lett. \textbf{92}, 133202 (2004).
\bibitem{remark_on_impossible_potentials_bis}
It can be shown for example that if the energy shift induced
by a single $\delta$-function scatterer situated in between
two walls is reinterpreted in terms of the interaction
present in the fermionic Hamiltonian (2), the second
order of the Taylor expansion in the powers of its prefactor,
proportional to $1/g_{\mbox{\scriptsize 1D}}$, turns out to be positive,
contradicting the non-positivity of the second order perturbation
theory shift; see Problem 4.1.11 in \cite{olshanii_34}.
\bibitem{girardeau2004_023608}
M.\ D.\ Girardeau and M.\ Olshanii, Phys. Rev. A \textbf{70}, 023608 (2004).
\bibitem{girardeau2003_0309396}
M.\ Girardeau and M.\ Olshanii, Preprint at arXiv:cond-mat/0309396 (2003).
\bibitem{diptiman2003_7517}
Diptiman Sen, J. Phys. A {\bf 36}, 7517 (2003).
\bibitem{lieb1963_1605}
E.\ H.\ Lieb and W.\ Liniger, Phys. Rev. \textbf{130}, 1605 (1963).
\bibitem{kohn1961_1242}
W.\ Kohn, Phys. Rev. \textbf{123}, 1242 (1961).
\bibitem{busch1998_549}
T.\ Busch, B.\ G.\ Englert, K.\ Rzazewski, and M.\ Wilkens, Found. Phys. \textbf{28}, 549 (1998).
\bibitem{astrakharchik2005_063620}
G.\ E.\ Astrakharchik, Phys. Rev. A \textbf{72}, 063620 (2005).
\bibitem{pitaevskii1998_4541}
L.\ P.\ Pitaevskii and S.\ Stringari, Phys. Rev. Lett. \textbf{81},4541 (1998).
\bibitem{landau_hydrodynamics}
L.\ D.\ Landau and E.\ Lifshitz, {\it Fluid Mechanics: v. 6 (Course of Theoretical Physics)} (Butterworth-Heinemann, Oxford, 1987)
\bibitem{griffin2009_book}
Allan Griffin, Tetsuro Nikuni, Eugene Zaremba, {\it Bose-Condensed Gases at Finite Temperatures} (Cambridge University Press, Cambridge, 2009)
\bibitem{regal2003_053201}
C.\ A.\ Regal, C.\ Ticknor, J.\ L.\ Bohn, and D.\ S.\ Jin, Phys. Rev. Lett. \textbf{90}, 053201 (2003).
\bibitem{gaebler2007_200403}
J.\ P.\ Gaebler, J.\ T.\ Stewart, J.\ L.\ Bohn, and D.\ S.\ Jin, Phys. Rev. Lett. \textbf{98}, 200403 (2007).
\bibitem{olshanii_34}
M.\ Olshanii, {\it Back-of-the-Envelope Quantum Mechanics: With Extension to Many-Body Systems and Integral PDEs} (Word Scientific, Singapore, 2013)
\bibitem{paraah2010_065603}
F.\ N.\ C.\ Paraan and V.\ E.\ Korepin, Phys. Rev. A \textbf{82}, 065603 (2010).
\bibitem{kohn1961_1242}
W.\ Kohn, Phys. Rev. \textbf{123}, 1242 (1961).
\bibitem{clark2010}
F.\ W.\ J.\ Olver, D.\ W.\ Lozier, R.\ F.\ Boisvert, and C.\ W.\ Clark, {\it NIST Handbook of Mathematical Functions} (Cambridge University Press 2010).
\bibitem{gradshteyn1997}
I.\ S.\ Gradshteyn and I.\ M.\ Ryzhik, {\it Table of Integrals, Series, and Products}, 7-th edition (Elsevier Academic
Press, Oxford, 2007).
\end{thebibliography}

\appendix
\section{}
The ground state of the unperturbed system is $|\Psi_{0}\rangle=\left(\prod_{n=0}^{N-1} \hat{b}_n^{\dagger}\right)|\mbox{vac}\rangle$ where $|\mbox{vac}\rangle$ is the
vacuum with no particles at all. The first order perturbation theory correction to ground state 
energy $E_0^{(0)}$ is given by 
\begin{equation}
\begin{split}
E_0^{(1)}=\frac{\hbar\omega}{\sqrt{\pi N}} & \frac{1}{\gamma_0(N)}\sum_{m=1}^{N-1}\sum_{n=0}^{m-1}\frac{(m-n)^2
\Gamma\left(m-\frac{1}{2}\right)}{\Gamma\left(m+1\right)}\\[0.2cm]
& \times\frac{\Gamma\left(n-\frac{1}{2}\right)}{\Gamma\left(n+1\right)}\ {}_3 F_2\left[\begin{matrix}\frac{3}{2},-n,-m\\[0.08cm]
\frac{3}{2}-n,\frac{3}{2}-m\end{matrix};1\right]
\end{split}
\label{6}
\end{equation}
and ${}_3 F_2\left[a_1,\,a_2,\,a_3;\,
b_1,\,b_2;\,z\right]$ is the order $(3,\,2)$ generalized hypergeometric function.
This result, as well as its derivation, is essentially identical to the formula obtained in
Ref.~\cite{paraah2010_065603}.

\section{}
There is only one eigenstate, $|\Psi_{1}\rangle=\hat{b}_N^{\dagger}\hat{b}_{N-1}^{}|\Psi_{0}\rangle$, 
in the first excited state manifold, and thus the correction to the energy is  
$E_1^{(1)} = \langle\Psi_{1}|\hat{\mathcal{V}}|\Psi_{1}\rangle$. The formula for the transition 
frequency, $\hbar\omega_{1,\,0} \equiv E_1-E_0$ , assumes a compact form, and it reads 
\begin{equation}
\begin{split}
\hbar\omega_{1,\,0}\equiv 
\hbar\Omega_{\mbox{\scriptsize D}}=\hbar\omega + {\cal O}(\frac{1}{\gamma_{0}(N)^2})
\end{split}
\end{equation}
We interpret the state $|\Psi_{1}\rangle$ as the first state of an infinite $\hbar\omega$-spaced ``dipole'' ladder: coherent wave 
packets formed out of the members of the ladder represent finite-amplitude dipole excitations (\textit{i.e.} oscillations of 
the center of mass); their frequency 
$\Omega_{\mbox{\scriptsize D}}$ is equal to the frequency of the trap \textit{exactly}, interactions notwithstanding
\cite{kohn1961_1242}. The zeroth state of the ladder is the
ground state.

\section{}
\subsection{Evaluation of $I_N^{(2a)}-I_N^{(2b)}$}
By using the identity for the generalized hypergeometric function \cite{clark2010}, one has
\begin{widetext}
\begin{equation}
\begin{aligned}
I_N^{(2a)}-I_N^{(2b)} & =\sqrt{\frac{32}{\pi^3}}\frac{\Gamma\left(N-\frac{1}{2}\right)\Gamma\left(N-\frac{5}{2}\right)}{\Gamma(N+1)\Gamma(N-1)}\bigg\{ {}_3 F_2\left[\begin{matrix}\frac{3}{2},2-N,-N\\[0.08cm] \frac{7}{2}-N,\frac{3}{2}-N\end{matrix};1\right]-\frac{\left(N-\frac{1}{2}\right)\left(N-\frac{5}{2}\right)}{(N+1)(N-1)}{}_3 F_2\left[\begin{matrix}\frac{3}{2},1-N,-N-1\\[0.08cm] \frac{5}{2}-N,\frac{1}{2}-N\end{matrix};1\right]\bigg\}\\[0.24cm]
& =\sqrt{\frac{72}{\pi^3}}\frac{\Gamma\left(N-\frac{5}{2}\right)\left(N+\frac{1}{2}\right)}{\Gamma(N)\Gamma(N+2)}{}_3 F_2\left[\begin{matrix}\frac{3}{2},1-N,-N\\[0.08cm] \frac{7}{2}-N,\frac{1}{2}-N\end{matrix};1\right]
\end{aligned}
\label{10}
\end{equation}
\end{widetext}

\subsection{Derivation of $I_N^{(2a)}+I_N^{(2b)}-2I_N^{(0)}$}
In addition, we also need to evaluate $I_N^{(2a)}+I_N^{(2b)}-2I_N^{(0)}$ and thus carry out the following calculation
\begin{equation}
\begin{split}
I_N^{(2a)}-\sum_{m=1}^{N-1}\sum_{n=0}^{m-1}\upsilon_{nm}=\upsilon_{N-1,N}-\upsilon_{N-1,N+1}
\end{split}
\label{12}
\end{equation}

\begin{equation}
\begin{split}
I_N^{(2b)}-\sum_{m=1}^{N-1}\sum_{n=0}^{m-1}\upsilon_{nm}=\upsilon_{N-1,N}-\upsilon_{N-2,N}
\end{split}
\label{13}
\end{equation}
and after some manipulations
\begin{widetext}
\begin{equation}
\begin{aligned}
\sqrt{\frac{\pi^3}{2}} \frac{N!(N-1)!}{\Gamma\left(N-\frac{1}{2}\right)\Gamma\left(N-\frac{3}{2}\right)} (\upsilon_{N-1,N+1}-\upsilon_{N-1,N})=\frac{3\left(N-\frac{1}{2}\right)(N-1)}{\left(N-\frac{5}{2}\right)(N+1)}{}_3 F_2\left[\begin{matrix}\frac{3}{2},-N,1-N\\[0.08cm] \frac{7}{2}-N,\frac{1}{2}-N\end{matrix};1\right]
\end{aligned}
\label{14}
\end{equation}
\end{widetext}
which leads to
\begin{equation}
\begin{split}
I_N^{(2a)}+I_N^{(2b)}-2I_N^{(0)}
& =-\sqrt{\frac{72}{\pi^3}}\frac{N\Gamma\left(N-\frac{5}{2}\right)\Gamma\left(N+\frac{1}{2}\right)}{\Gamma(N)\Gamma(N+2)}\\[0.2cm]
& \qquad\times {}_3 F_2\left[\begin{matrix}\frac{3}{2},1-N,-N\\[0.08cm] \frac{7}{2}-N,\frac{1}{2}-N\end{matrix};1\right]
\end{split}
\label{16}
\end{equation}
where $I_N^{(0)}$ and $\upsilon_{nm}$ have been already given before


\subsection{Derivation of $\Omega_N$}
We define $R_{mn}(x)=\frac{1}{2}\left[\varphi'_m(x)\varphi_n(x)-\varphi_m(x)\varphi'_n(x)\right]$, then it can be shown
\begin{equation}
\begin{split}
R_{mn}(x)= & \frac{m-n}{\sqrt{\pi 2^{m+n}n!m!}}\sum_{k=0}^m 2^k k!\begin{pmatrix}m\\k\end{pmatrix}\begin{pmatrix}n\\k\end{pmatrix}\\[0.2cm]
& \qquad\quad\times e^{-x^2}H_{m+n-2k-1}(x)
\label{17}
\end{split}
\end{equation}
where $m<n$, $\varphi_n$ is the $n$-th eigenfunction of harmonic trap and $H_n(x)$ is the $n$-th Hermite polynomial. Then we have
\begin{widetext}
\begin{equation}
\begin{aligned}
& \int_{-\infty}^{+\infty}R_{N-2,N+1}R_{N-1,N}dx\\[0.2cm]
& =\frac{3}{\pi(N-2)!N!\sqrt{N^2-1}}\frac{1}{2\sqrt{2}}\sum_{k=0}^{N-2}\sum_{l=0}^{N-1}(-1)^{k+l}k!l!\begin{pmatrix}N-2\\k\end{pmatrix}\begin{pmatrix}N+1\\k\end{pmatrix}\begin{pmatrix}N-1\\l\end{pmatrix}\begin{pmatrix}N\\l\end{pmatrix}\Gamma\left(2N-k-l-\frac{3}{2}\right)\\[0.2cm]
& =-\frac{3}{8}\sqrt{\frac{2(N^2-1)}{\pi^3}}\frac{\Gamma\left(N-\frac{5}{2}\right)\Gamma\left(N+\frac{1}{2}\right)}{\Gamma(N)\Gamma(N+2)}{}_3 F_2\left[\begin{matrix}\frac{3}{2},1-N,-N\\[0.08cm] \frac{7}{2}-N,\frac{1}{2}-N\end{matrix};1\right]
\end{aligned}
\label{18}
\end{equation}
\end{widetext}
Thus $\Omega_N$ is given
\begin{widetext}
\begin{equation}
\begin{aligned}
\Omega_N =-8\int_{-\infty}^{+\infty}R_{N-2,N+1}R_{N-1,N}dx=\sqrt{\frac{18(N^2-1)}{\pi^3}}\frac{\Gamma\left(N-\frac{5}{2}\right)\Gamma\left(N+\frac{1}{2}\right)}{\Gamma(N)\Gamma(N+2)}{}_3 F_2\left[\begin{matrix}\frac{3}{2},1-N,-N\\[0.08cm] \frac{7}{2}-N,\frac{1}{2}-N\end{matrix};1\right]
\end{aligned}
\label{19}
\end{equation}
\end{widetext}
where the formula for integral \cite{gradshteyn1997}
\begin{equation}
\begin{aligned}
\int_{-\infty}^{+\infty} e^{-2x^2}H_p(x)H_q(x)dx= & (-1)^{(p+3q)/2}2^{(p+q-1)/2}\\
& \times\Gamma\left(\frac{p+q+1}{2}\right)
\end{aligned}
\end{equation}
was used. The monopole frequency (3) and (4) are immediately arrived from these results.

\end{document}